\documentclass[twocolumn]{aastex62}
\usepackage{amsmath}

\newcommand{\bz}{$\langle B_z \rangle$}

\received{----}
\revised{----}
\accepted{----}
\submitjournal{ApJ}

\shorttitle{ECME from HD 142990}
\shortauthors{Das et al.}

\begin{document}

\title{Detection of coherent emission from the Bp star HD\,142990 at uGMRT frequencies}

\correspondingauthor{Barnali Das}
\email{barnali@ncra.tifr.res.in}

\author[0000-0001-8704-1822]{Barnali Das}
\affil{National Centre for Radio Astrophysics, Tata Institute of Fundamental Research,  Pune University Campus, Pune-411007, India}

\author[0000-0002-0844-6563]{Poonam Chandra}
\affil{National Centre for Radio Astrophysics, Tata Institute of Fundamental Research,  Pune University Campus, Pune-411007, India}

\author{M. E. Shultz}
\affiliation{Annie Jump Cannon fellow, Department of Physics and Astronomy, University of Delaware, 217 Sharp Lab, Newark, DE 19716, USA}

\author{Gregg A. Wade}
\affiliation{Department of Physics, Royal Military College of Canada, PO Box 17000, Station Forces, Kingston, ON K7K 7B4, Canada}


\begin{abstract}
HD\,142990 is a Bp-type star with a nearly dipolar surface magnetic field of kG strength. 
Recently \cite{lenc18} reported the tentative discovery of Electron Cyclotron Maser Emission (ECME) from this star at 200 MHz. This type of emission has been observed from only three other hot magnetic stars. In this paper, we present our observations of HD\,142990 with the upgraded Giant Metrewave Radio telescope (uGMRT) at 550-804 MHz and with the legacy GMRT at 1420 MHz near the rotational phases of the nulls of the longitudinal magnetic field. We found strong enhancements in flux density in both circular polarisations suggesting an ECME bandwith of at least 1.2 GHz (200-1420 MHz). In one of the observation sessions, we observed enhancements with opposite circular polarisations from the same magnetic pole. This has not been reported in any other hot magnetic star known to exhibit ECME. In order to explain this unusual finding, we suggest a scenario that involves a transition of the dominant mode of ECME between the magneto-ionic modes.
\end{abstract}


\section{Introduction} \label{sec:intro}
Roughly 10\% of early type (OBA) stars are found to have strong ordered (mostly dipolar) surface magnetic fields \citep{grunhut17,sikora18}. Non-thermal radio emission is expected to arise from such stars due to the acceleration of charged particles in the stellar wind in the presence of a magnetic field. \cite{drake87} first detected such emission from five Ap/Bp stars with the Very large Array (VLA). 
From the observed brightness temperature and the degree of circular polarisation, the emission mechanism was inferred to be gyrosynchrotron \citep{drake87,linsky92}. The strength of the radio emission was found to be rotationally modulated, and this modulation correlates with that of the longitudinal magnetic field \citep[e.g.][]{leone93,lim96} indicating that the emission arises near the polar regions \citep{linsky92,leone93}.

In addition to gyrosynchrotron emission, a small number of hot magnetic stars have been discovered to emit intense, periodic pulsed emission with a very high degree ($\sim$100\%) of circular polarisation. These pulses always arrive near the rotational phases at which the stellar longitudinal magnetic field $\langle B_z \rangle$   is zero. This phenomenon was first discovered in the Ap star CU Vir \citep{trigilio00}. Based on the pulse characteristics, the authors suggested that it was caused by Electron Cyclotron Maser Emission (ECME). This was later confirmed by \cite{trigilio08} who showed that the plasma frequency at the region where ECME originates is much smaller than the electron gyrofrequency. \cite{leto16} proposed a tridimensional model to simulate the ECME pulse profile from a star with an ideal axi-symmetric dipolar field. According to this model, for a CU Vir-like star (orientation of the line of sight and the magnetic field axis w.r.t. the rotation axis similar to that of CU Vir), there will be two pairs of observable pulses over one rotation cycle. Each pair consists of one Left Circularly Polarised (LCP) pulse and one Right Circularly Polarised (RCP) pulse situated on either side of the rotational phase corresponding to a magnetic null (which we will refer to as a magnetic null phase). Also, if for the first pair, the LCP pulse arrives before the RCP pulse, for the other pair, it is the RCP pulse that arrives first followed by the LCP pulse. However, no LCP pulse from CU Vir has yet been reported. The same is the case with the Bp star HD\,133880, which is the second hot magnetic star from which ECME has been observed \citep{chandra15,das18}. Both LCP and RCP pulses were first observed from the star HD\,142301 \citep{leto19}. The observation of the LCP pulse not only boosts confidence in the current understanding of the ECME mechanism, but also enabled the authors to infer the  magneto-ionic mode of ECME, which turned out to be the ordinary mode.

In addition to CU Vir, HD\,133880 and HD\,142301, there is one more hot magnetic star suggested to display ECME. Highly circularly polarised emission was observed from HD\,142990 by \cite{lenc18} during an all sky circular polarisation survey with the Murchison Widefield Array (MWA) at 200 MHz. Unlike the other stars with ECME, the variation of the flux density with rotational phase of the star was not available and as a result, ECME could not be confirmed. However, using the fact that ECME is active in hot magnetic stars and that it gives rise to highly circularly polarised emission, the authors suggested HD\,142990 to be a tentative host of ECME. 

In this paper, we present the ECME pulse profile for the star HD\,142990 obtained with the upgraded Giant Metrewave Radio Telescope (uGMRT) in the frequency range of 550-804 MHz (band 4). Our observations not only confirm ECME, but also for the first time suggest the signature of a transition between the magneto-ionic modes. In addition, we also present the result of our observations of this star with the legacy GMRT (GMRT before its upgrade) at 610 MHz and 1420 MHz and with a bandwidth of 33.33 MHz.


This paper is structured as follows: in the next section (\S\ref{sec:ecme}), we briefly review the ECME phenomenon in the context of hot magnetic stars. After that we give an introduction to the star HD\,142990 (\S \ref{sec:hd142990}), which is followed by the observation and data analysis section (\S \ref{sec:data}). This is followed by the results (\S\ref{sec:results}) and discussions (\S \ref{sec:discussion}). We summarise our findings in the conclusion (\S \ref{sec:conc}). 

\section{Electron Cyclotron Maser Emission}\label{sec:ecme}
\begin{figure}
\centering
\includegraphics*[width=0.47\textwidth]{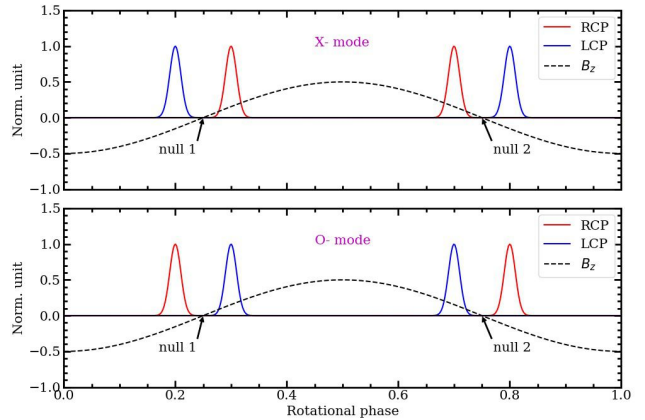}
\caption{A cartoon diagram showing ECME from a star with an axi-symmetric dipolar magnetic field with two observable magnetic nulls. The nearly constant base flux has been neglected. In the upper panel, the expected ECME pattern for X-mode emission is shown. In the lower panel, the same is shown for O-mode emission. The dotted curve in each panel represents the longitudinal magnetic field averaged over the stellar disk \bz.\label{fig:mock_ecme}}
\end{figure}

\begin{figure*}
\centering
\includegraphics*[trim={0cm 2cm 0cm 2.5cm}, clip, scale=0.7]{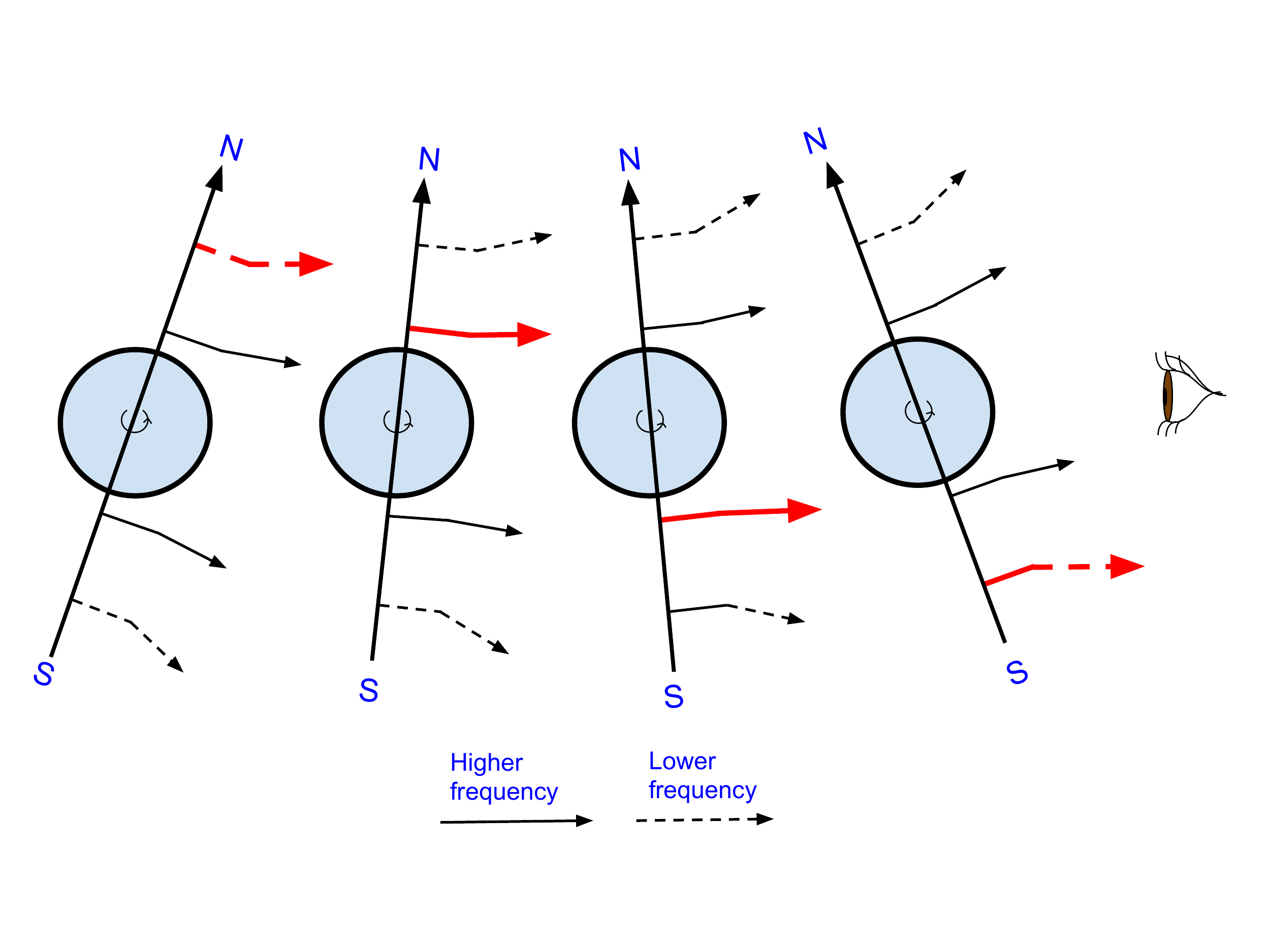}
\caption{A cartoon diagram illustrating how the frequence dependence of pulse arrival time arises by the diferent amounts of deviation suffered by pulses at different frequencies. The solid arrows represent the higher frequency pulses (originate closer to the star) and the dashed arrows represent lower frequency pulses (originate farther from the star); the line of sight is in the horizontal direction. The star is rotating counter-clockwise. Whenever a  pulse comes to the field of view, it is shown with a red arrow. See \S \ref{sec:ecme} for details. \label{fig:freq_depend_model}}
\end{figure*}
Electron Cyclotron Maser Emission or ECME is a type of coherent emission that occurs in a magnetised plasma with an anisotropic particle distribution. The emission occurs at frequencies close to the harmonics of the electron-cyclotron frequency. For a mildly relativistic electron population, the direction of emission is almost perpendicular to the local magnetic field line \citep{melrose82}. In order to explain the highly directed nature of ECME from hot magnetic stars, \cite{trigilio11} suggested the `tangent plane beaming model', according to which the maser amplification occurs tangentially to annular rings above the magnetic poles so that the direction of emission is perpendicular to the local magnetic field line and also parallel to the magnetic equatorial plane.

ECME is highly circularly polarised \citep{melrose82} and the handedness of the circular polarisation (right circular or left circular) depends on the magneto-ionic mode of emission. For $\omega_\mathrm{p}/\omega_\mathrm{B}<<1$ (where $\omega_\mathrm{p}$ and $\omega_\mathrm{B}$ are respectively the plasma frequency and the electron gyrofrequency), the extra-ordinary (X-) mode is preferred over the ordinary (O-) mode; for $0.3-0.35<\omega_{\rm p}/\omega_{\rm B}\leq 1 $, the O-mode overtakes the X-mode \citep{melrose84,sharma84,leto19}. In the case of X-mode emission, the sense of rotation of the electric field vector is the same as the helicity of the emitting electrons, whereas in O-mode, it is the opposite \citep{trigilio00,leto19}. 

In hot magnetic stars, electrons gyrate anti-clockwise near the North pole and clockwise near the South pole \citep{trigilio00}. Thus for X-mode emission, we expect to see right circularly polarised (RCP) radiation from the North pole and left circularly polarised (LCP) radiation from the South pole. For ECME in the O-mode, the situation is the opposite. This is illustrated in Figure \ref{fig:mock_ecme} which is a cartoon illustrating the relation between the longitudinal magnetic field curve and the polarisation of the radio pulses. As mentioned in the preceding paragraph, the pulses are directed nearly perpendicular to the magnetic field which results in the fact that they are seen only near the rotational phases where the longitudinal magnetic field averaged over the stellar disk \bz~is zero. From this point onwards, we will refer to the rotational phase where \bz~changes from negative to positive as null 1 and to that where \bz~changes from positive to negative as null 2. Near null 1, the South pole is receding and the North pole is approaching; we hence expect to see the pulse(s) from the South pole first followed by the pulse(s) from the North pole. By symmetry, near null 2, we expect to see the pulse(s) from the North pole first followed by the pulse(s) from the South pole \citep{leto16}. Thus, for X-mode emission, we first expect to see LCP followed by RCP near null 1 and RCP followed by LCP near null 2 (upper panel of Figure \ref{fig:mock_ecme}), whereas for O-mode emission, we expect the opposite sequence of pulse-arrival (lower panel of Figure \ref{fig:mock_ecme}). In this way the magneto-ionic mode of ECME can be determined, provided that \bz~is known, which can then be used to constrain $\omega_\mathrm{p}$ \citep[e.g.,][]{leto19}.

One important aspect of ECME is that ECME pulses at different frequencies do not arrive at the same rotational phases \citep[e.g.][]{trigilio11,lo12,leto16}. As discussed by \cite{leto16}, the observed frequency dependence can arise by two mechanisms: difference in intrinsic angle of emission, and refraction experienced while passing through the inner magnetosphere. 
\cite{lo12} assumed that the pulses are intrinsically emitted perpendicular to the field axis irrespective of the frequency of emission, and the observed frequency-dependence arises solely due to propagation effects \citep[this framework was first proposed by ][]{trigilio11}. With this model they successfully reproduced the pulse arrival sequence for CU\,Vir at two frequencies: 1.0 GHz and 2.3 GHz . This implies that over this frequency range (fractional bandwidth=79\%), the effect of differences in the intrinsic angle of emission is negligible compared to that due to propagation effects. 


The effect of propagation through the inner magnetosphere is to refract the wave upwards \citep[e.g. see Figure 2 of][]{leto16}. This deviation is greater for lower frequencies than for higher frequencies, which is a consequence of the fact that the refractive index inside the inner magnetosphere increases with frequency \citep[\S 3.5 and Figure 10 of][]{lo12}.

In Figure \ref{fig:freq_depend_model}, we have illustrated how this fact gives rise to a frequency-dependent pulse arrival time. We have taken the simple case where the line of sight (LoS) and the magnetic dipole axis are perpendicular to the axis of rotation. We have also assumed rigid rotation for simplicity. The circle in each of the four sub-figures represents the star and the arrow passing through the circle represent the direction of the magnetic field axis. The LoS is to the right (shown by an eye) and in the horizontal direction. The four sub-figures show four phases of the star as \bz~passes through zero from positive to negative, i.e., the North pole of the star is receding from the observer and the South pole is approaching the observer. In each sub-figure, the ECME pulses at two frequencies originating near the two poles are shown. The higher frequency (shown with solid arrows) originates closer to the star than the lower frequency (shown with dashed arrows). As described in the preceding paragraph, the deviation of the lower frequency emission is greater than that of the higher frequency (assuming the original direction to be perpendicular to the magnetic field axis). 
A pulse is received by an observer only when the direction of the corresponding arrow becomes parallel to the line-of-sight (in this case horizontal). In the first sub-figure, the lower frequency pulse from the North pole (red dashed arrow) becomes visible to the observer as its direction becomes horizontal. In the next phase (second sub-figure), the higher frequency pulse from the same pole (red solid arrow) is received by the observer. After that \bz~becomes negative and the pulses from the South pole come into view. At first, the higher frequency pulse (from the South pole, shown by the red solid arrow in the third sub-figure) enters the line-of-sight which is followed by the lower frequency pulse (red dashed arrow in the fourth sub-figure). The net result is that when \bz~changes from positive to negative, the expected pulse-arrival sequence is the following: a lower frequency pulse from the North pole, a higher frequency pulse from the same pole, a higher frequency pulse from the South pole and finally a lower frequency pulse from the same pole. Note that the arrival sequences of the lower and higher frequency pulses are opposite for the two magnetic poles. 


\section{HD 142990}\label{sec:hd142990}

\begin{deluxetable}{lll}
\tablecaption{Stellar and magnetic properties of HD\,142990  \citep{shultz16,shultz18,shultz19_1} and Shultz et al. (in prep.) \label{tab:hd142990}} 
\tablehead{
\hline
Physical quantity & Value & Unit
}
\startdata
\hline
Mass $(M)$ & 5.7$\pm$ 0.1 & $M_\odot$\\
Radius $(R_*)$ &  2.80$\pm$ 0.04 & $R_\odot$ \\
 Kepler radius $(R_\mathrm{K})$ &  2.58$\pm$0.03 & $R_*$ \\
Alfv\'en radius ($R_\mathrm{A}$) &  22.1$\pm$0.8 & $R_*$ \\
Temperature $(T_\mathrm{eff})$ & 18.0$\pm$ 0.5 & kK \\
Luminosity $(\log(L/L_\odot))$ & 2.9$\pm$0.1 & -\\
Dipole strength $(B_\mathrm{d})$ & 4.7$\pm$0.3 & kG \\
Surface gravity $(\log\mathrm{g})$ & 4.15$\pm$ 0.11 & -\\
Inclination angle\tablenotemark{a} $(i)$ & 54.6$\pm$1.5 & degree\\
Obliquity\tablenotemark{b} $(\beta)$ &  89.84$^{+0.03}_{-3.01}$ & degree\\
Fractional age\tablenotemark{c} $(\tau_\mathrm{MS})$ &  0.28$^{+0.10}_{-0.07}$ & -\\
\hline
\enddata
\tablenotetext{a}{Angle between the rotation axis and the line of sight}
\tablenotetext{b}{Angle between the rotation axis and the magnetic field axis}
\tablenotetext{c}{Ratio of the current age of the star to the total time for which the star will be in the main-sequence (MS), calculated using the evolutionary models of \cite{ekstroem12}}
\end{deluxetable}

HD\,142990 is a chemically peculiar (He weak) B5 V type star \citep{houk88} located in the Scorpius constellation. The values of some of the physical parameters for this star are given in Table \ref{tab:hd142990}. The longitudinal magnetic field ($B_{\rm z}$) of this star has been found to vary between $\approx \pm 1$ kG with rotational phase and this variation can be best fitted by a second-order sinusoid \citep{shultz18}. This indicates that the magnetic field geometry of this star is likely not purely dipolar.

\cite{petit13} classfied the magnetosphere of HD\,142990 as a centrifugal magnetosphere based on the value of its Kepler radius ($R_\mathrm{K}$) and Alfv\'en radius ($R_\mathrm{A}$). 
The Alfv\'en radius indicates the distance from the star at which the magnetic field still dominates the wind \citep{ud-doula02}, and the Kepler corotation radius is the distance from the star at which the centrifugal force acting upon the corotating plasma equals the gravitational force \citep{townsend05,ud-doula09}. Since $R_\mathrm{A}> R_\mathrm{K}$, the star has a `centrifugal magnetosphere', a warped disk within which plasma is able to accumulate over long timescales \citep{townsend05}. \cite{petit13} noted that stars with $R_\mathrm{A}>> R_\mathrm{K}$ tend to show variable $\mathrm{H_\alpha}$ emission consistent with the pattern predicted by the Rigidly Rotating Magnetosphere model advanced by \cite{townsend05}. HD\,142990 has indeed been reported to display variable $\mathrm{H_\alpha}$ emission, as well as emission from ultraviolet resonance lines, believed to originate in its magnetosphere \citep{shore04}.


The observational record of HD 142990 spans many decades. Time-series photometric observations were first reported by \cite{pedersen77}. \cite{borra83} reported the first detection of the star's magnetic field and demonstrated that the magnetic measurements could be phased  with a period of $\approx$ 0.98 d or one-half of that value. Modern studies of the photometric, magnetic, and spectroscopic variability of HD 142990 (e.g. Shultz et al. 2018 and submitted, Bowman et al. 2018) demonstrate that the variability of the star is monoperiodic with a period of $\approx$ 0.98 d and that the variability is consistent with the rotational modulation of surface chemical abundance inhomogeneities, as is typical for CP stars.

Although both \cite{shultz18} and \cite{bowman18} obtained a stellar rotation period close to 0.98 d, the two measurements are not in agreement within the quoted uncertainties. The rotation period of \cite{shultz18}, 0.978832(2) d is smaller than the value 0.97892(2)d reported by \cite{bowman18} by $\approx$ 7.6s. Investigation of this discrepancy led Shultz et al. (submitted) to discover that the period is not constant but decreasing at a rate of $\approx$ 0.58$\pm$ 0.01 s/yr. 
In this paper, we have adopted this variable rotation period ephemeris. Using $\mathrm{dP/dt=\dot{P}}=-0.58$ s/yr, and taking the reference HJD ($\mathrm{HJD_0}$) and rotation period ($\mathrm{P_0}$) as 2442820.93 and 0.979110 d respectively (Shultz et al. submitted), we get the following analytical form for the ephemeris in the limit $\mathrm{(\dot{P}/P_0)}\Delta t<<1$ to calculate the rotational phases:

\begin{align}\label{eq:ephem}
E&=\frac{2\Delta t}{\mathrm{2P_0}+\mathrm{\dot{P}}\Delta t}
\end{align}
Where $\Delta t=\mathrm{HJD-HJD_0}$. The rotational phases corresponding to the magnetic nulls (0.250 and 0.664) were obtained via a harmonic fit to \bz~phased using Eq. \ref{eq:ephem}.

\section{Observations and data analysis}\label{sec:data}
We observed HD\,142990 with the uGMRT on 2018 July 27, 2018 August 13 and 2018 August 23 in band 4 (550-850 MHz). The original bandwidth of observation on 2018 July 27 and 2018 August 13 was 200 MHz (550-750 MHz) and that on 2018 August 23 was 400 MHz (550-950 MHz). The difference in bandwidth arises because of an upgrade of the observatory (the data on 23 August were acquired through a DDT proposal after the upgrade).
Observations were scheduled in order to cover the rotational phases near the magnetic null phases of the star. 
Data for both the circular polarisations, $RR$ (Stokes I+V) and $LL$ (Stokes I-V), were recorded. Stokes $RR$ corresponds to RCP and Stokes $LL$ corresponds to LCP. The standard flux calibrator 3C286 was observed at the beginning and/or at the end of each observing session to calibrate the absolute flux scale. Each scan of the target (HD\,142990) was preceded and followed by the observation of a phase calibrator: J1626-298 for the observations on 13 August and 23 August and J1626-298 and J1517-243 for the observations on 27 July. The on-source time varied from 2--2.7 hours in various observation sessions. 
\par 
The data were analysed using the Common Astronomy Software Applications package \citep[\textsc{casa},][]{mcmullin07}. Dead antennae were flagged by manual inspection using the tasks `plotms' and `flagdata'. Most of the Radio Frequency Intereference (RFI) was removed by running the automatic flagging algorithm `rflag' incorporated into the task `flagdata'. Any remaining corrupted data were then flagged manually. The edges of the band were also flagged due to very low gain. The final widths of the bands were 166 MHz (560-726 MHz, on 27 July and 13 August) and 234 MHz (570-804 MHz, on 23 August).

In order to improve the bandpass calibration (obtaining the frequency-dependent part of the antenna gains), we used the phase calibrator(s), in addition to the flux calibrator, as bandpass calibrators. This can be done only if the spectral indices of the calibrators are known. For that, the flux densities of the phase calibrators were obtained at different locations of the band from the known flux of 3C286 using the task `fluxscale'. The spectral index for J1517-243 came out to be zero within the errorbars, whereas that for J1626-298 came out to be -0.3$\pm$0.1 within the observing band. The model visibilities were then created using the task `setjy'. After that, we ran bandpass (task `bandpass') with both flux and phase calibrators to obtain the frequency-dependent antenna gains. The time-dependent parts were obtained using the task `gaincal'. These calibrations were applied to all the sources and the corrected data were inspected for bad data which were then flagged using the task `flagdata'. After that the data were recalibrated. This flagging+calibration cycle was repeated until the corrected data did not show any significant presence of RFI.

The calibrated data for HD\,142990 were then averaged over a few frequency channels. The final spectral resolution was 0.58 MHz (this value was chosen so as to avoid bandwidth smearing). The next step was imaging and self-calibration. We used the tasks `clean' and `gaincal' for this purpose. In order to improve the image quality, we used an offline flagging routing `ankflag' (Bera in prep.) which identifies and removes any remaining RFI from the residuals obtained after subtracting out the model visibilities from the self-calibrated data. We tried to obtain at least one measurement of flux density per scan ($\approx$40 minutes). However, in some cases, we had to average over a longer timerange so as to improve the significance of the detection. For scans with enhanced flux densities, we imaged every 4 minutes of data and self-calibration was done for each such time slice. This strategy was applied for RCP and LCP independently.

As mentioned in \S \ref{sec:intro}, we had also observed the star with the legacy GMRT at 1420 MHz and at 610 MHz. The bandwidth in each case was 33.3 MHz. The observation at 1420 MHz was carried out on 2014 November 8, and the one at 610 MHz was carried out on 2015 September 10. The analysis of these data was done in a manner similar to that described above.

\begin{figure*}
\centering
\includegraphics*[width=0.95\textwidth]{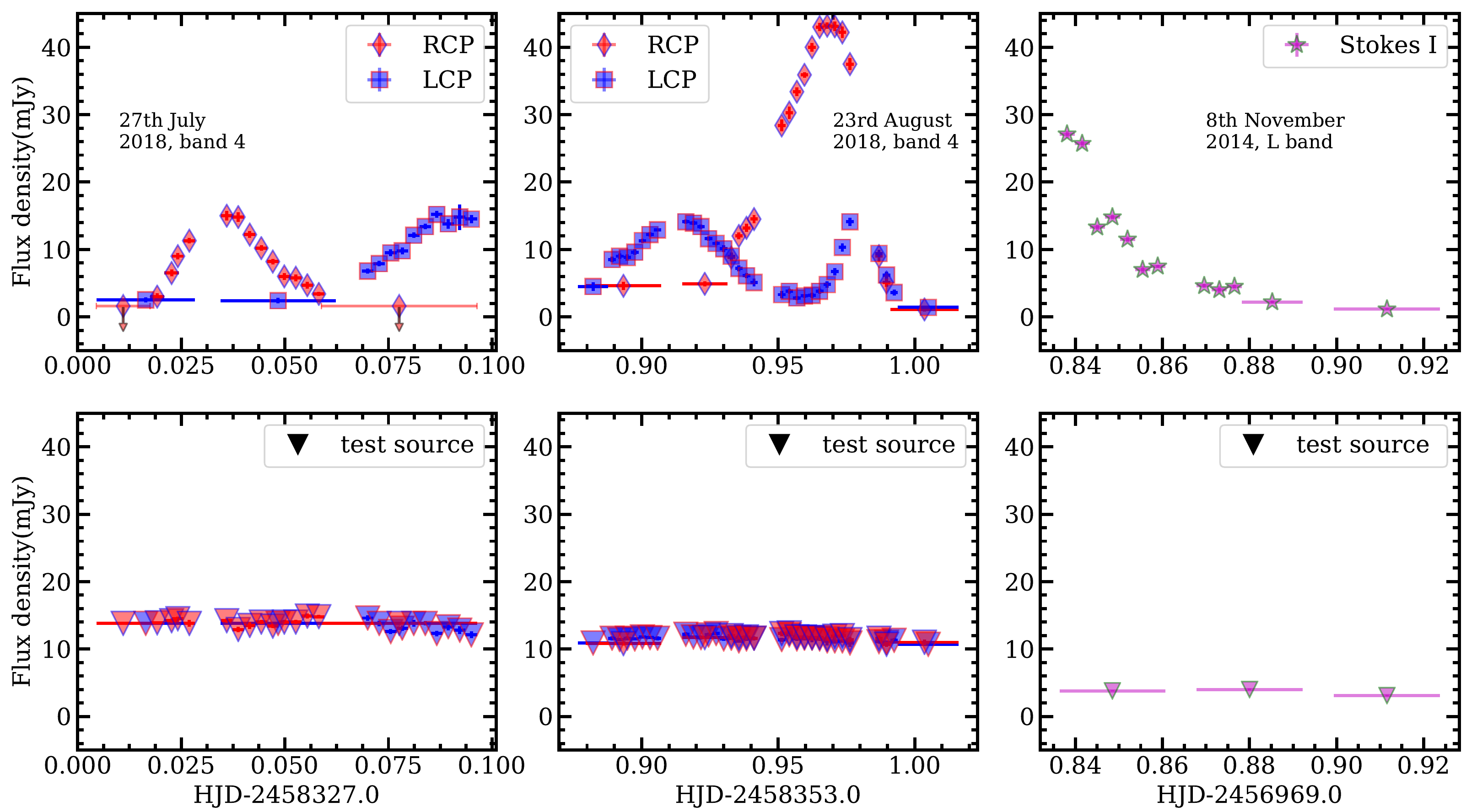}
\caption{\textit{Upper panels}: The lightcurves of HD\,142990 in band 4 (first and second columns) and L band (third column) near its magnetic nulls. The red diamonds are for RCP (Stokes $RR$), blue squares are for LCP (Stokes $LL$) and the stars in magneta represent the average of RCP and LCP (Stokes I). \textit{Lower panels}: The time-variation of the flux density of a test source in the field of view on the three days. Red and blue indicate RCP and LCP respectively and the magneta points are for Stokes I. The flux densities are fairly constant meaning that the observed flux density variations for HD\,142990 are real.  \label{fig:ecme_lightcurves}}
\end{figure*}





\begin{figure*}
\centering
\includegraphics*[width=0.9\textwidth]{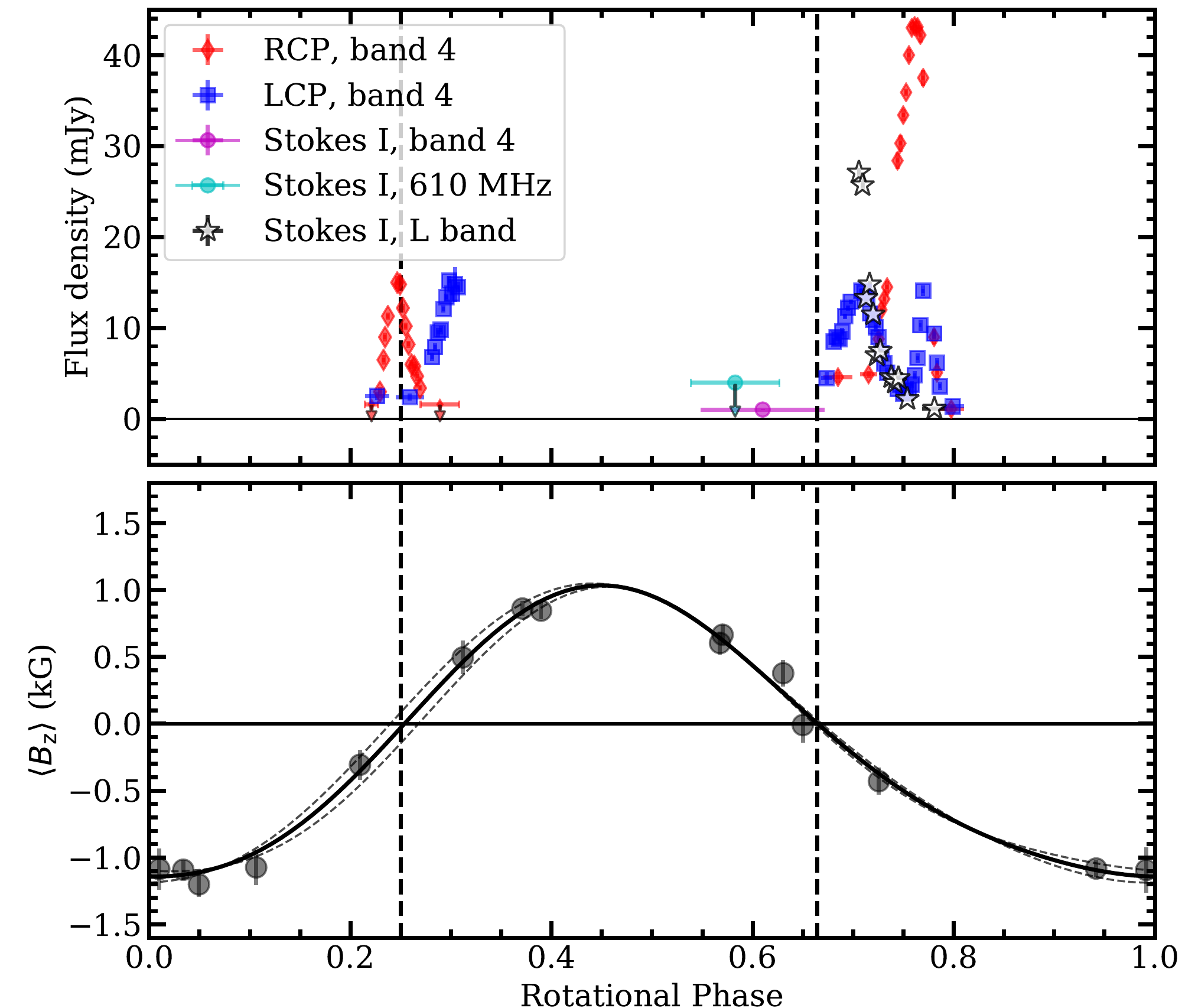}
\caption{\textit{Upper panel:} The data acquired in 2018 (band 4, uGMRT), 2015 (610 MHz, legacy GMRT) and 2014 (1420 MHz or L band, legacy GMRT) phased with the ephemeris of  Shultz et al. (submitted). The purple and cyan circles correspond to data taken on 2018 August 13 in band 4 and on 2015 September 10 at 610 MHz respectively. The first set of pulses in band 4 were observed on 2018 July 27 and the second set of pulses in band 4 were observed on 2018 August 23. The stars represent the data taken on 2014 November 8 in L band. \textit{Lower panel:} The ESPaDOnS H line \bz~measurements \citep[reported by][]{shultz18} of the star, phased with the same ephemeris (black circles). The solid and dashed curves represent the best second-order harmonic fit and the $1\sigma$ uncertainty respectively. The dashed vertical lines correspond to the magnetic null phases.\label{fig:linear_var_ephem}}
\end{figure*}

\section{Results}\label{sec:results}

The flux densities of HD\,142990 for the data taken on 2018 July 27, 2018 August 23, and those acquired on 2014 November 8, are given in Tables \ref{tab:27jul_data}, \ref{tab:23aug_data} and \ref{tab:L_band_data} respectively. 
For the data taken on 2018 August 13 (in band 4), we had to average over all the scans and both polarizations in order to be able to detect the star. The flux density obtained at this epoch was 1.04$\pm$0.08 mJy over the HJD range 2458343.97146 - 2458344.09206. For the data taken on 2015 September 10, the star wasnot detected even after averaging over all the scans and polarisations. The 4$\sigma$ upper limit to the flux density (Stokes I) came out to be 4.0 mJy over the HJD range of 2457276.05359-2457276.13989.

We observed flux enhancements in both band 4 and L band (Figure \ref{fig:ecme_lightcurves}). 
The data were then phased using Eq. \ref{eq:ephem}. In Figure \ref{fig:linear_var_ephem}, we show the phased radio data (upper panel) as well as ESPaDOnS H line \bz~measurements reported by \cite{shultz18}. 
The enhancements observed on 2018 July 27 occur close to null 1 and those on 2018 August 23, as well as at 1420 MHz (L band), occur close to null 2. 
We attribute these enhancements to ECME for the following reasons:
\begin{enumerate}
\item Enhancements occur near both the magnetic nulls, although there are offsets ($\approx 0.02$ near null 1 and $\approx 0.06$ and near null 2). Offsets of ECME pulses \citep[as large as 0.07,][]{kochukhov14} from the magnetic null phases have also been observed for two of the other three hot magnetic stars exhibiting ECME, namely CU Vir \citep{trigilio00,kochukhov14} and HD\,142301 \citep{leto19}. The offsets can arise due to a differential rotation between the photosphere and the radio emitting regions \citep{pyper13}, or a magnetic field topology more complex than that of an axi-symmetric dipole \citep{leto19}, or both.

\item The enhancements are highly circularly polarised ($>50\%$) \footnote{The observed circular polarisation is not 100\% because of the overlap of the RCP and LCP pulses. The 100\% circular polarisation can be observed only if the pulses are well-separated \citep{leto16}. } and highly directional in all cases.

\item The occurrence of pulses near null 2 in both the 2014 and 2018 data indicates the phenomenon persistently recurs at similar rotation phases, rather than being a transient phenomenon that was coincidentally observed at the expected rotational phase.
\end{enumerate}
In the following subsections, we describe the key features of the ECME observed on 2018 July 27, 2018 August 23, and 2014 November 8.

\subsection{ECME pulses observed on 2018 July 27}\label{subsec:ecme_27jul}
The variation of the flux density with rotational phase observed on 2018 July 27 is shown in the upper left panel of Figure \ref{fig:ecme_lightcurves}. As predicted by the ECME model of \cite{leto16}, there are enhancements in both RCP (red diamonds) and LCP (blue squares). The maximum observed circular polarisations of the RCP and the LCP pulses are 72\% and 80\% respectively. 
If we now consider the fact that the observed flux densities include a `base flux density' due to gyrosynchrotron which has a very low circular polarisation near the magnetic nulls \citep{lim96, das18}, we can exclude this base flux density while calculating the percentage circular polarisation for the observed pulses. Once we do that, the maximum observed circular polarisations for the RCP and the LCP pulses increase to 82\% and 92\% respectively. For the base flux density, we have used the flux density obtained from our observation on 2018 August 13 which is $\approx$ 1 mJy. 


As explained already in \S \ref{sec:ecme} and Figure \ref{fig:mock_ecme}, it is possible to infer the magneto-ionic mode of emission if we can observe both RCP and LCP pulses of ECME. In this case, the enhancements occur close to null 1, where \bz~passes through zero from negative to positive values (see Figure \ref{fig:linear_var_ephem}). Combining this observation with the fact that we observed the RCP pulse first followed by the LCP pulse, we can conclude that the emission corresponds to O-mode \citep{leto19}.

\subsection{ECME pulses observed on 2018 August 23}\label{subsec:ecme_23aug}

\begin{figure*}
\centering
\includegraphics*[width=0.9\textwidth]{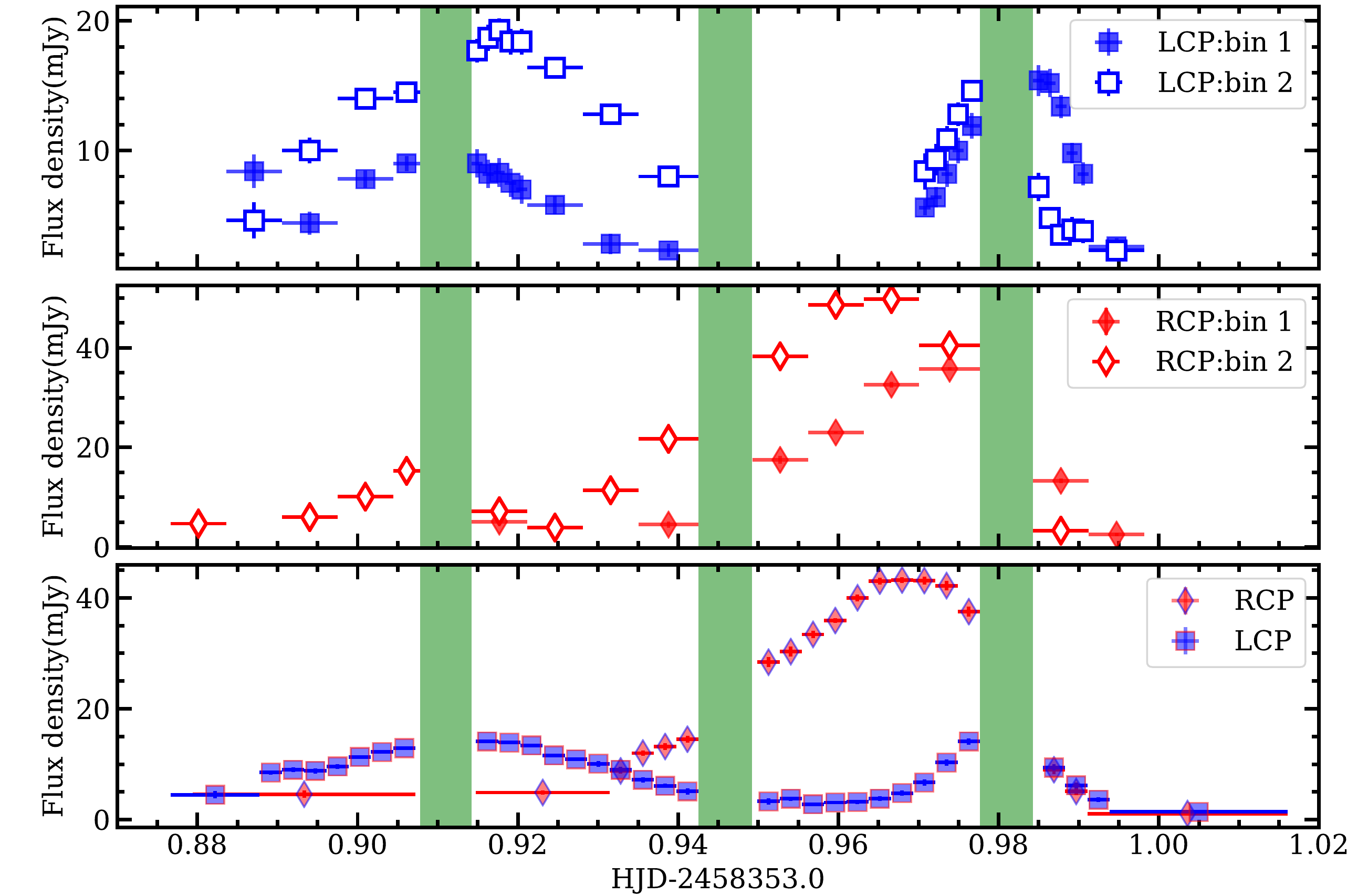}
\caption{\textit{Upper and middle panel}: Variation of flux density on 2018 August 23 with rotational phases for LCP (blue points) and RCP (red points) for the two frequency bins. Bin 1 represents the frequency range 569.8-686.4 MHz and bin 2 represents the frequency range 687-803.6 MHz. The filled markers are for the lower frequency bin (bin 1) and the unfilled ones are for the higher frequency bin (bin 2). The green columns mark phase-gaps due to the observation of the phase calibrator. \textit{Bottom panel}: The variation of flux density with rotational phases for the full band. Red and blue points are for RCP and LCP respectively. \label{fig:freq_binning}}
\end{figure*}

The result of our observation of HD 142990 on 2018 August 23 is shown in the upper right panel of Figure \ref{fig:ecme_lightcurves}. Unlike the pulses observed on 27 July, the shapes of the pulses observed on this day deviate significantly from those expected from a star with an ideal dipolar magnetic field. The amplitudes of the enhancements in the two polarisation are significantly different.  
The maximum observed circular polarisations for the LCP and the RCP pulses are 48\% and 80\% respectively, and increase to 54\% and 84\% respectively once we exclude the base flux from the calculation.

The most striking feature that we observed on this day is the double-peaked LCP pulse. The observation of the second peak of the LCP pulse made it difficult to infer the mode of emission of the pulses. For emission in O-mode (as found for our observation on 27 July), we should obtain only the first peak of the LCP pulse followed by the RCP pulse since the enhancements occur close to null 2.

We attempted to understand the reason behind this peculiar pulse arrival sequence which was seen only on this date (near null 2). The primary difference between the data obtained on 23 August and 27 July is their observing frequency ranges: whereas the 27 July observation spans the frequency range 560-726 MHz (166 MHz bandwidth), the 23 August observation spans the frequency range 570-804 MHz (234 MHz bandwidth). This prompted us to check if the peculiarity in our 23 August observation is a result of averaging over the broader range of frequencies. For this purpose, the frequency range of observation on 23 August was divided into two frequency bins of equal width. Bin 1 covers the frequency range 569.8-686.4 MHz and bin 2 covers the frequency range 687.0-803.6 MHz. 
We obtained the lightcurves for each bin for the two circular polarisations. The presence of a very bright extended source close to the star (approximately 2 arcminutes away) made this task very difficult as the noise is higher near a bright source compared to those at locations far away from bright sources. As a result, we could detect the star only at rotational phases during which there were enhancements. The result of this exercise is shown in Figure \ref{fig:freq_binning} (LCP in the upper panel and RCP in the middle panel). For comparison, we also show the pulses obtained for the full band in the bottom panel. The filled markers represent bin 1 (the lower frequency bin) and the unfilled markers represent bin 2 (the higher frequency bin). The green shaded regions correspond to phase gaps due to the observation of the phase calibrator(s). Other regions devoid of any points signify non-detection of the star at those phases. 

Below, we describe the LCP pulse-profile (\S \ref{subsubsec:lcp_freq_binning}) and the RCP pulse-profile (\S \ref{subsubsec:rcp_freq_binning}) and the information that can be extracted from them.

\subsubsection{Variation of the LCP pulse-profile with frequency}\label{subsubsec:lcp_freq_binning}
From the upper panel of Figure \ref{fig:freq_binning} we see that the sequences in which pulses at the two frequency bins arrive are opposite for the two peaks. Since the maximum fractional bandwidth in band 4 is only 37\%, it is justified to assume that the difference in pulse-arrival times within band 4 arises primarily due to the propagation effects. In that case, the two peaks of the LCP pulse cannot originate near the same magnetic pole (\S \ref{sec:ecme}).

At null 2, \bz~changes its sign from positive to negative. This implies that, near this phase, the pulse originating near the North pole should arrive first, followed by the one from the South pole (\S \ref{sec:ecme}). In our case, we found that the two LCP peaks are actually two separate LCP pulses coming from two different magnetic poles. We can therefore associate the first peak of the LCP pulse with the North pole and the second peak with the South pole. Now, from \S \ref{sec:ecme}, we expect that near null 2, for the pulses coming from the North pole, the lower frequency will arrive before the higher frequency and for the ones from the South pole, the higher frequency will arrive before the lower frequency. This is what we see here which provides additional support to the interpretation that the first LCP peak originates near the North pole and the second one originates near the South pole.

\subsubsection{Variation of the RCP pulse-profile with frequency}\label{subsubsec:rcp_freq_binning}
In the middle panel of figure \ref{fig:freq_binning}, we show the RCP pulses observed on 2018 August 23 for the two frequency bins. While only one RCP peak was visible in the full band (bottom panel of Figure \ref{fig:freq_binning}), dividing it into two bins revealed that there is another RCP peak at the higher frequency bin that arrives before the strong RCP peak.
This additional pulse, though much weaker than the main RCP pulse, is comparable in amplitude to the LCP pulses observed on the same day. Since this pulse is detected for only one of the frequency bins, it is unclear whether it is a part of the stronger RCP pulse at the same frequency or has a completely different site/mechanism of origin. For the stronger RCP pulse, it is evident that the pulse at the higher frequency arrives before the one at the lower frequency. This sequence is the same as that for the second peak of the LCP pulse observed on 23 August and we conclude that the second peak of the LCP pulse and the strong RCP pulse observed on 23 August both originate near the South pole. This association of the strong RCP pulse with the South pole is consistent with the fact that it is observed after the first LCP peak which has been associated with the North pole.

In \S \ref{sec:discussion}, we discuss the possible mode(s) of emission suggested by the observed pulses. 

\subsection{L band observation}\label{subsec:L_band_obs}

\begin{figure*}
\centering
\includegraphics*[width=0.95\textwidth]{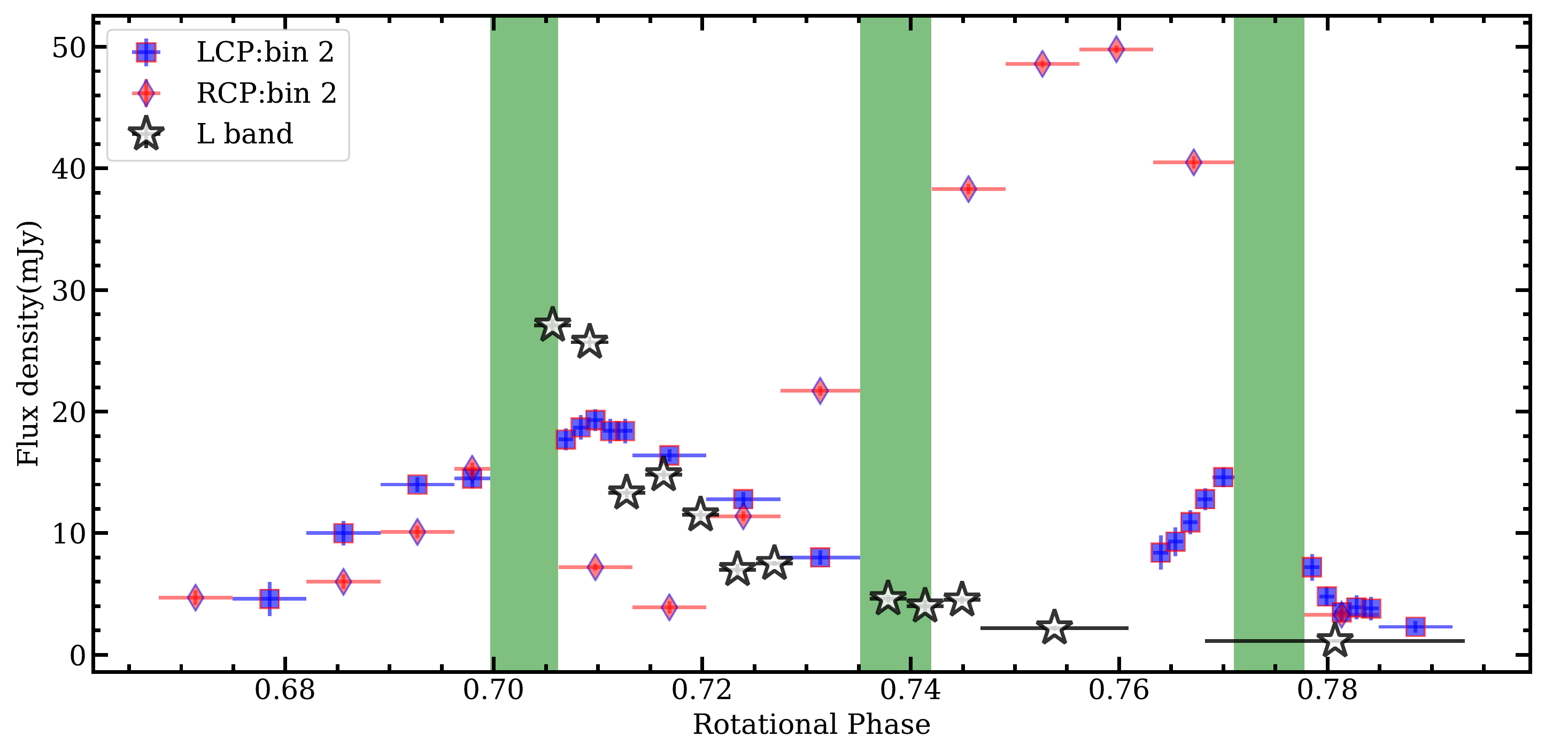}
\caption{Observed lightcurve at 1420 MHz (L band) for total intensity (purple stars), data were acquired on 2014 November 8. For comparison, we have also plotted the lightcurves obtained on 2018 August 23 for bin 2 (687-804 MHz) of band 4 (cyan circles). The nearest magnetic null lies at the rotational phase 0.66 (null 2).\label{fig:L_band_lightcurve}}
\end{figure*}

The L band (1420 MHz) data were obtained on 2018 November 8 with the GMRT. The rotational phase range covered by these data is close to null 2 (similar to the data obtained on 2018 August 23). Since the cross-correlation outputs in GMRT L band correspond to linear polarisations and there was no polarisation observation, we could retrieve the lightcurve only for the total intensity (Stokes $I$) which is the average of RCP and LCP. 

In Figure \ref{fig:L_band_lightcurve}, we plot the L band lightcurve on top of the band 4 lightcurves for bin 2 (687-803.6 MHz) obtained from our observation on 2018 August 23. Our data covered only the descending portion of the ECME pulse-pair at L band. Therefore, the observed portion of the enhancement is likely to be due to the pulse coming from the South pole, which is the one that arrives after the arrival of the pulse from the North pole (for null 2). Now, from \S \ref{sec:ecme}, we know that near null 2, because the L band enhancement is coming from the South pole, it is expected to lie after the pulse from the North pole irrespective of the pulse frequency \footnote{This is valid even when we do not make the assumption that the differences in pulse-arrival times are primarily due to propagation effect. This is because, the intrinsic angle of emission w.r.t. the magnetic field is $\leq 90^o$. Therefore in all cases we expect to see the pulses from the North pole first followed by the ones from the South pole, if \bz~is changing from positive to negative values.}. This is clearly not the case in Figure \ref{fig:L_band_lightcurve}, since the maximum observed flux density in L band lies before the peak of the first LCP pulse associated with the North pole. Thus the relative phase shifts observed between the L band data taken in 2014 and band 4 data taken in 2018 are not expected from our current understanding. This is further discussed in \S \ref{subsec:new_ephemeris} .



\section{Discussion}\label{sec:discussion}
From our observations of HD\,142990 with the uGMRT, we inferred that ECME pulses of opposite circular polarisations can arise from the same magnetic poles. To the best of our knowledge, this phenomenon has not been observed for any other hot magnetic star. In the next subsection (\S \ref{subsec:ecme_mode}), we discuss a scenario which can give rise to this phenomenon. Another important result is that while comparing the ECME pulses from two different epochs, we found that there is a relative phase shift between the pulses which is inconsistent with the current understanding of the behaviour of ECME pulses at different frequencies. In \S\ref{subsec:new_ephemeris}, we discuss the implication of this finding.
\subsection{The origin of the doubly peaked LCP pulse: a consequence of mode-transition?}\label{subsec:ecme_mode}
From the sequence of arrival of LCP and RCP pulses, we deduced that the mode of emission for the ECME pulses observed on 2018 July 27 is the ordinary mode. However, presence of two LCP peaks near null 2 on 23 August made it difficult to deduce the same inference.


As mentioned in \S\ref{sec:ecme}, the dominant magneto-ionic mode of ECME is the X-mode for $\omega_\mathrm{p}/\omega_\mathrm{B}\leq 0.3-0.35$, and O-mode for $0.3-0.35<\omega_\mathrm{p}/\omega_\mathrm{B}\leq 1$ \citep{melrose84,sharma84,leto19}. The frequency of emission is proportional to the local gyrofrequency and consequently higher frequencies arise closer to the star than the lower frequencies.  Thus, as we observe ECME at lower frequencies, we are probing the magnetosphere farther away from the star. With increasing distance from the star, the values of $\omega_\mathrm{p}$ and $\omega_\mathrm{B}$ decrease as $r^{-\frac{3}{2}}$ and $r^{-3}$ respectively \citep{trigilio08}, where $r$ is the radial distance from the star. Consequenctly, even if for higher frequencies (closer to the stellar surface), the condition for X-mode emission is satisfied, there will be a frequency below which the condition for O-mode emission will be satisfied and the mode of emission will make a transition from X-mode to O-mode. 

This transition is expected to be continuous rather than an abrupt change of mode. A consequence of this is that there will exist a frequency range over which emissions in both modes will be visible. We suggest that the frequency range in band 4 corresponds to this transition frequency range, where we are observing ECME in both O-mode and X-mode. If we associate the weak RCP enhancement (middle panel of Figure \ref{fig:freq_binning}), detected only for the higher frequency bin, with the North pole, it supports the picture of the presence of an X-mode component, the strength of which is steeply decreasing with decreasing frequency.


At this point, the question arises as to why we saw the X-mode component from the South pole throughout the observing band on 23 August and did not see it on 27 July. A probable answer to this question is that ECME growth rates are highly variable with time \citep[ECME pulse amplitude variation within a short timescale has also been observed for CU Vir,][]{trigilio11}. From the upper panel of Figure \ref{fig:ecme_lightcurves}, we see that on 27 July, the pulse heights are the same for both polarisations, however on 23 August, the RCP pulse is much stronger than the LCP pulse. This indicates that on 27 July, the ECME growth rates are comparable at both magnetic poles, but on 23 August, there was an increase in the growth rates at the South pole (at which the strong RCP pulse originates), which could also increase the growth rate for X-mode emission making it detectable throughout the band. 


\begin{figure}
\centering
\includegraphics[width=0.48\textwidth]{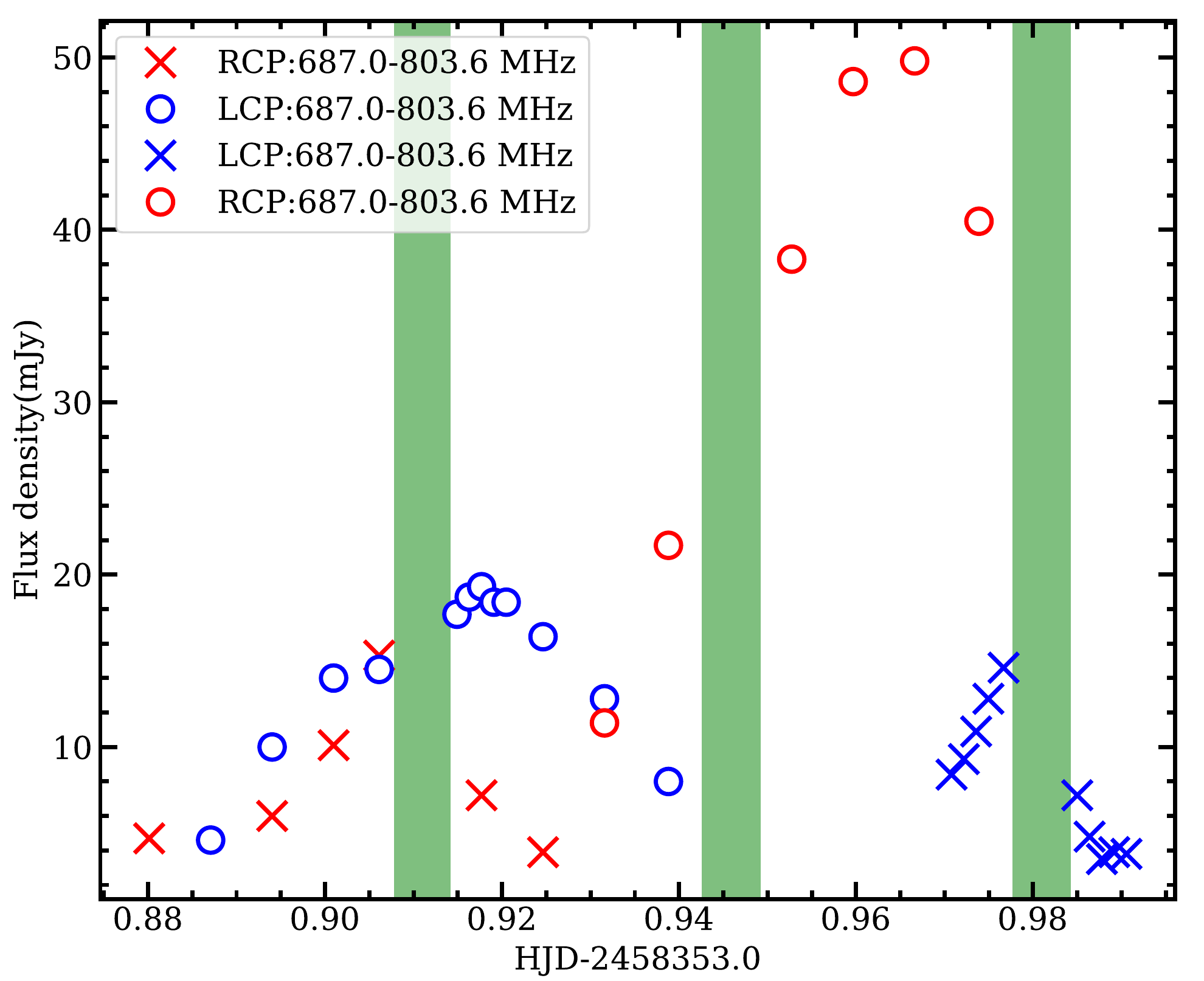}
\caption{The proposed picture of ECME at X and O-modes. These data are for the higher frequency bin obtained after dividing the full band of observation on 2018 August 23 into two equal sub-bands. The blue points are for LCP and the red points are for RCP. The LCP and RCP pulses are assigned to either O-mode (marked with circles) or X-mode (marked with `x's) emission based on our hypothesis (refer to \S \ref{subsec:ecme_mode}). The resultant picture is shown here. At first, we receive the X-mode component from the North pole (red `x's), followed by the O-mode component from the same pole (blue circles). After that, we receive the O-mode component from the South pole (red circles) which is followed by the X-mode component from the same pole (blue `x's). Errorbars are omitted in this figure to enhance clarity. The green columns represent phase gaps due to the observation of the phase calibrator.\label{fig:X_O_mode}}
\end{figure}

Another fact that supports this picture of mode transition is the sequence of arrival times of the pulses that we are suggesting to be X-mode components. The refractive index of the X-mode component is lower than that for the O-mode component at the same frequency. Consequently, the X-mode component will suffer a larger deviation while passing through the inner magnetosphere than that for the O-mode component. This is similar to the case that for a given mode of emission, the lower-frequency pulses experience larger deviations than those at the higher frequencies since the former have lower refractive indices (\S \ref{sec:ecme}). In our case, if we consider only the pulses (near null 2) associated with the O-mode emission (the first peak of the LCP pulse and the strong RCP pulse shown in the upper and middle panels of Figure \ref{fig:freq_binning} respectively), we get the following sequence of arrival for the pulses in the two frequency bins: a lower-frequency component (LCP), then a higher-frequency component (LCP), followed by a higher-frequency component (RCP) and finally a lower-frequency component (RCP).
Mapping the lower frequency component to X-mode and the higher frequency to O-mode, we would expect to receive first an X-mode component, which would be followed by an O-mode component, then another O-mode component and finally another X-mode component. This is exactly what we have observed (see  Figure \ref{fig:X_O_mode}) making the picture internally consistent.

We would like to clarify that the above picture of transition from one magneto-ionic mode to another is suggestive at this stage and needs further evidence. One caveat of this picture is that the association of the weak RCP pulse (middle panel of Figure \ref{fig:freq_binning}) to the same pole as the one at which the first peak of the LCP pulse (upper panel of Figure \ref{fig:freq_binning}) originates is arbitrary. 
Besides, the difference in amplitudes between the RCP and LCP pulses from the South pole is much larger than that between the oppositely polarised pulses from the North pole (see Figure \ref{fig:X_O_mode}) which means that the physical conditions near the two magnetic poles are not the same. This is not a surprise given that the star's magnetic field is likely to have a more complex structure than that of an axi-symmetric dipole \citep{shultz18}. Thus the picture of mode transition, though suggestive at this stage, can explain all the observed features of the ECME pulses
and hence is a very plausible explanation. An immediate prediction of this hypothesis is that at a sufficiently high frequency, but lower than the upper cut-off for ECME, the mode of emission should be entirely X-mode, and at a sufficiently low frequency, but higher than the lower cut-off for ECME, the mode of emission should be entirely O-mode. For example, if we assume that the ratio $\nu_{\rm p}/\nu_{\rm B}=0.3$ for the higher frequency bin ($\nu_{\rm avg}=745.3$ MHz), the ECME pulse arrival sequence (w.r.t. RCP and LCP) should be opposite at 2.3 GHz and 333 MHz if our hypothesis is correct. While saying this, we are making the assumption that the ECME cut-offs for this star are outside the frequency range of 0.33-2.3 GHz. 

If the frequency $\nu_{\rm avg}=745.3$ MHz (the average frequency for bin 2) is indeed the transition frequency, it implies a plasma frequency of $\nu_{\rm p}\approx 223.6$ MHz for emission at the fundamental harmonic and $\nu_{\rm p}\approx 117.8$ MHz for emission at the second harmonic at the region of emission. The corresponding plasma density at the region of emission is then $\approx 6\times10^8$ cm$^{-3}$ (fundamental harmonic) and $\approx 2\times 10^8$ cm$^{-3}$ (second harmonic). Note that \cite{leto19} estimated a plasma density of $10^9-10^{10}$ cm$^{-3}$ for HD\,142301 at the region where ECME at 1.5 GHz originates. For $B_d\approx$ 14 kG \citep{leto19}, 1.5 GHz emission at the fundamental harmonic corresponds to a height of $\approx$ 2 $R_* \equiv 5$ $R_\odot$ from the surface of HD\,142301 \citep[$R_*=2.52$ $R_\odot$,][]{kochukhov06}. In the case of HD\,142990, emission at 745 MHz at the fundamental harmonic corresponds to a height of 1.4 $R_* \equiv 4.1$ $R_\odot$ from the stellar surface. Thus our estimation of the plasma density for HD\,142990's magnetosphere is at least an order of magnitude lower than that estimated for HD\,142301's at a similar height. This suggests that HD\,142990 has a faster wind \citep[since number density goes as $r^{-3}/v_\mathrm{wind}$,][]{trigilio08}. However, further study is needed to confirm the density estimate as well as to find out the wind speeds around the two stars.

\begin{figure}\label{fig:P_dot_changed}
\centering
\includegraphics*[width=0.45\textwidth]{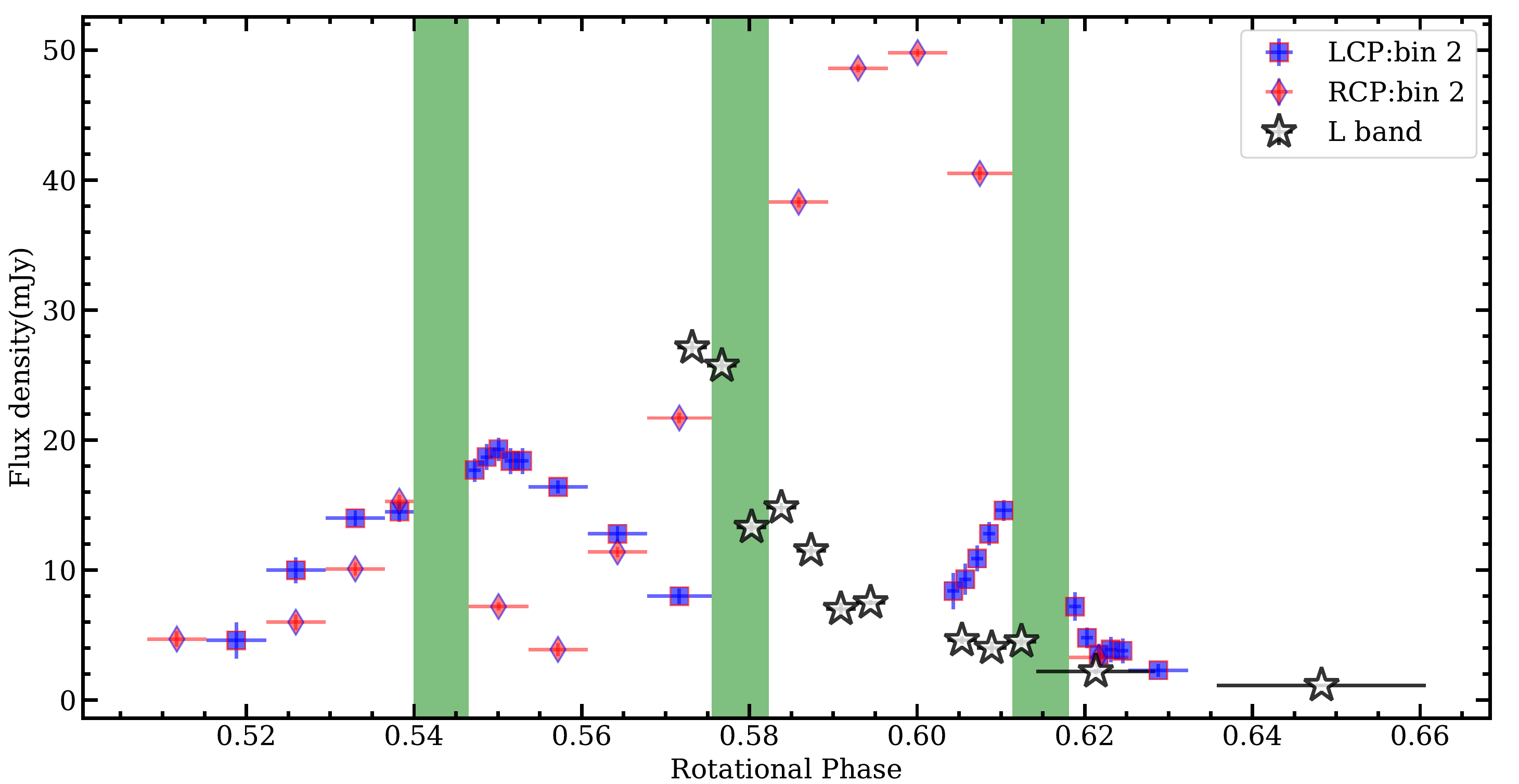}
\caption{Same as Figure \ref{fig:L_band_lightcurve}, but assuming $\mathrm{\dot{P}}=-0.54$ s/yr (instead of -0.58 s/yr) while phasing the data. \label{fig:band4_L_band_lower_P_dot}}
\end{figure}

\subsection{A non-uniformly changing rotation period?}\label{subsec:new_ephemeris}
ECME pulses observed in 2014 (in L band) and in 2018 (in band 4) cannot be phased consistently using the linearly decreasing rotation period model of Shultz et al. (submitted), if we associate the observed enhancement in L band with the South magnetic pole of the star. One way to resolve this problem is to assume that the L band enhancement came from the North pole and the pulse from the South pole was very weak on that day. Given that the variability in ECME pulse strength has already been observed in CU\,Vir \citep{ravi10,trigilio11}, we cannot rule out the same for HD\,142990. Future observation in L band with circular polarisation information and a better rotational phase coverage will be useful to find out the phases of arrival of the L band pulses relative to that in band 4.

Another possibility is that the L band enhancement indeed came from the South pole, but the ephemeris is incorrect. Shultz et al. (submitted) obtained the variable period ephemeris by using data spanning the years 1981--2015. Thus while phasing the band 4 data, we have extrapolated this model. We found that a lower $\mathrm{\dot{P}}$, such as   -0.54 s/yr, improves the consistency between the L band and the band 4 data as shown in Figure \ref{fig:P_dot_changed}. With this value of $\mathrm{\dot{P}}$, the offsets of the ECME pulses (band 4) from the magnetic nulls came out to be $\approx$ 0.02 (null 1) and $\approx$ 0.04 (null 2). The consistency can also be obtained with $\mathrm{\dot{P}}=0$ if we use the rotation period obtained from the ESPaDOnS data alone, which is 0.97887(6) d (Shultz et al. submitted). The corresponding offsets of the band 4 pulses from the magnetic nulls are $\approx$ 0.03 (null 1) and $\approx$ 0.025 (null 2) if we use the same reference HJD as is used in Eq. \ref{eq:ephem}. These conclusions are qualitatively consistent with the oscillatory period model briefly discussed by Shultz et al. (submitted). Note that the rotation period of the star CU\,Vir has also been found to show an oscillatory evolution with time \citep{mikulasek11}.


The observed discrepancy in the radio data indicates the need to obtain new measurements in order to further constrain the evolution of the star's rotation period. On the other hand, the rotation period change implied by the radio data may have nothing to do with the photospheric rotation period. \cite{pyper13} reported that certain period glitches indicated by radio data of the star CU\,Vir, are not consistent with photospheric measurements which led them to suggest that the radio emitting region and the photosphere might not be in perfect co-rotation with each other. In both the cases, new measurements of the stellar rotation period (magnetic or photometric), as well as future radio data will be essential to understanding the real behaviour of the evolution of the stellar rotation period.

\section{Conclusions}\label{sec:conc}
HD\,142990 is the first hot star in which the magnetosphere is detectable in $\mathrm{H_\alpha}$, UV \citep{shore04} and ECME. In this paper, we have presented the its lightcurves in band 4 (550-804 MHz) and in L band (1420 MHz) near the magnetic null phases. The band 4 data were obtained with the uGMRT in the year 2018 whereas those in the L band were obtained with the legacy GMRT in the year 2014. We observed significant enhancements in flux density in both bands. In band 4, we found enhancements in both circular polarisations, which is a signature of ECME. This is only the fourth star from which ECME has been detected and only the second star, after HD\,142301 \citep{leto19}, for which ECME pulses with left circular polarisation have been observed. In L band, although we did not have any circular polarisation information, we suggest the enhancements to be of ECME origin since they are observed near the magnetic null phases. Combining it with the previous claim of ECME at 200 MHz \citep{lenc18}, we conclude that ECME is likely to be present at least in the frequency range of 200-1420 MHz.

We have also observed that the ECME pulses are slightly offset from their respective expected phases of arrival. Such offsets have also been observed in CU Vir \citep{kochukhov14} and in HD\,142301 \citep{leto19}. Phase offsets have been suggested to arise due to complex surface magnetic topologies \citep[e.g.][]{leto19}, or due to an imperfect corotation of the radio emitting region with the stellar photosphere \citep{pyper13}.
A magnetic field topology more complex than a perfect dipole has already been suggested for HD\,142990 by \cite{shultz18}. 

We also found that the phasing of the radio data obtained in 2014 and 2018 are not in agreement with each other unless we assume that on one of the days (in the L band data), the pulses from one of the magnetic poles were not detectable at all. Without this assumption, the phasing can be improved by using a slower value of $\mathrm{\dot{P}}$. Further radio data, as well as photometric/spectropolarimetric data are needed to understand the evolution of the stellar rotation period.

The most interesting result of this investigation is that on one of the days, we have observed pulses with opposite circular polarisations from the same magnetic pole. This is an unusual finding and to the best of our knowledge, has not been reported for any other star exhibiting ECME. To explain this observation, we  suggest that the frequency range of our observation is the one in which the dominant mode of emission changes from extra-ordinary to ordinary with decreasing frequency. Although the current data are not sufficient to accept/discard this hypothesis, future multifrequency observations with telescopes like the JVLA and the uGMRT will surely be able to unravel the true nature of ECME from this star.

\acknowledgements
We thank the referee for the useful comments which helped us to improve our manuscript. We would like to thank Dr. Emil Lenc for the useful informations provided by him regarding ECME from HD 142990 at 200 MHz. Our sincere gratitude will be to Prof. Yashwant Gupta for quickly approving our DDT proposal without which we would not have been able to complete this study.
PC  acknowledges support from the Department of Science and Technology via 
SwarnaJayanti Fellowship awards (DST/SJF/PSA-01/2014-15). GAW acknowledges Discovery Grant support from the Natural Sciences and Engineering Research Council (NSERC) of Canada. MS acknowledges support from the Annie Jump Cannon Fellowship, supported by the University of Delaware and endowed by the Mount Cuba Astronomical Observatory. We thank the staff of the GMRT that made these observations possible. 
The GMRT is run by the National Centre for Radio Astrophysics of the Tata Institute 
of Fundamental Research. This research has made use of NASA's Astrophysics Data System.

\appendix

\startlongtable
\begin{deluxetable*}{cc|cc}
\tablecaption{Variation of flux density of HD 142990 in band 4, data were acquired with the uGMRT on 2018 July 27\label{tab:27jul_data}} 
\tablehead{
\hline
Mean  & Flux density        & Mean & Flux density\\
HJD\tablenotemark{a}  & mJy & HJD\tablenotemark{a} & mJy
}
\startdata
\hline
LCP & & LCP\\
\hline
2458327.00452$\pm$0.01191 &  2.5$\pm$0.4 & 2458327.07971$\pm$0.00139  & 12.1$\pm$0.4\\
2458327.03449$\pm$0.013885  & 2.4$\pm$0.4 & 2458327.08249$\pm$0.00139  & 13.4$\pm$0.4\\
2458327.06859$\pm$0.001395 & 6.8$\pm$0.4 & 2458327.08527$\pm$0.00139  & 15.2$\pm$0.5\\
2458327.07138$\pm$0.00139  & 7.9$\pm$0.4 & 2458327.08805$\pm$0.001385  & 13.8$\pm$0.7\\
2458327.07416$\pm$0.00139  & 9.5$\pm$0.5 & 2458327.09082$\pm$0.00139  & 14.8$\pm$1.9\\
2458327.07694$\pm$0.001385  & 9.8$\pm$0.5 & 2458327.0936$\pm$0.00139  & 14.5$\pm$0.6\\
\hline
RCP & & RCP \\
\hline
2458327.01099$\pm$0.00647  & 1.6\tablenotemark{b} & 2458327.04431$\pm$0.00139  & 10.2$\pm$0.4\\
2458327.019195$\pm$0.001735  & 3.0$\pm$0.5 & 2458327.04709$\pm$0.00139  & 8.2$\pm$0.4\\
2458327.02267$\pm$0.00174  & 6.5$\pm$0.5 & 2458327.049865$\pm$0.001385  & 6.0$\pm$0.5\\
2458327.024175$\pm$0.001385  & 9.0$\pm$0.5 & 2458327.05264$\pm$0.00139  & 5.8$\pm$0.5\\
2458327.02695$\pm$0.00139  & 11.3$\pm$0.4 & 2458327.05542$\pm$0.00139  & 4.7$\pm$0.5 \\
2458327.03598$\pm$0.00139  & 15.0$\pm$0.7 & 2458327.0582$\pm$0.00139  & 3.4$\pm$0.3 \\
2458327.038755$\pm$0.001385  & 14.8$\pm$0.7 & 2458327.07759$\pm$0.01879  & 1.6\tablenotemark{b} \\
2458327.04153$\pm$0.00139  & 12.2$\pm$0.5 & - &-\\
\hline
\enddata
\tablenotetext{a}{The errors in the mean HJDs actually correspond to the length of the timerange over which the radio data were averaged}
\tablenotetext{b}{4$\sigma$ upper limit}
\end{deluxetable*}

\startlongtable
\begin{deluxetable*}{cc|cc}
\tablecaption{Variation of flux density of HD 142990 in band 4, data were acquired with the uGMRT on 2018 August 23\label{tab:23aug_data}} 
\tablehead{
\hline
Mean &Flux density          & Mean & Flux density \\
HJD\tablenotemark{a}  & mJy & HJD\tablenotemark{a} & mJy
}
\startdata
\hline
 LCP & & LCP\\
\hline
2458353.882215$\pm$0.00555  & 4.5$\pm$0.6 & 2458353.93838$\pm$0.00139  & 6.1$\pm$0.4\\ 
2458353.88916$\pm$ 0.00139  & 8.5$\pm$0.4 & 2458353.94116$\pm$0.00139  & 5.1$\pm$0.6\\
2458353.89194$\pm$0.00139  & 9.0$\pm$0.4 & 2458353.95128$\pm$0.00139  & 3.3$\pm$0.6\\
2458353.894715$\pm$0.001385  & 8.8$\pm$0.5 & 2458353.95406$\pm$0.00139  & 3.8$\pm$0.4\\
2458353.89749$\pm$0.00139  & 9.6$\pm$0.4 & 2458353.95684$\pm$0.00139 & 2.8$\pm$0.3\\
2458353.90027$\pm$0.00139  & 11.3$\pm$0.3 & 2458353.959615$\pm$0.001385  & 3.1$\pm$0.3\\
2458353.90305$\pm$0.00139  & 12.2$\pm$0.3 & 2458353.96239$\pm$0.00139  & 3.2$\pm$0.4\\
2458353.905825$\pm$0.001385  & 12.9$\pm$0.2 & 2458353.96517$\pm$0.00139  & 3.8$\pm$0.4\\
2458353.91616$\pm$0.00139  & 14.1$\pm$0.3 & 2458353.96795$\pm$0.00139  & 4.8$\pm$0.5\\
2458353.91894$\pm$0.00139  & 13.9$\pm$0.3 & 2458353.970725$\pm$0.001385  & 6.7$\pm$0.5\\
2458353.921715$\pm$0.001385 &13.4$\pm$0.3 & 2458353.9735$\pm$0.00139  & 10.3$\pm$0.6\\
2458353.92449$\pm$0.00139  &11.6$\pm$0.3 & 2458353.97628$\pm$0.00139  & 14.1$\pm$0.6 \\
2458353.92727$\pm$0.00139  &10.9$\pm$0.3 & 2458353.986905$\pm$0.001385  & 9.4$\pm$0.6\\
2458353.93005$\pm$0.00139  & 10.1$\pm$0.5 & 2458353.98968$\pm$0.00139  & 6.2$\pm$0.5\\
2458353.932825$\pm$0.001385  & 9.0$\pm$0.5 & 2458353.99246$\pm$0.00139  & 3.6$\pm$0.4\\
2458353.9356$\pm$0.00139  & 7.2$\pm$0.5 & 2458354.00496$\pm$0.011  & 1.4$\pm$0.2\\
\hline
RCP & & RCP \\
\hline
2458353.893325$\pm$0.013885  & 4.6$\pm$0.6 & 2458353.96239$\pm$0.00139  & 40.0$\pm$0.6\\
2458353.923105$\pm$0.008335  & 4.9$\pm$0.4 & 2458353.96517$\pm$0.00139  & 43.0$\pm$0.6\\
2458353.932825$\pm$0.001385  & 8.8$\pm$0.5 & 2458353.96795$\pm$0.00139  & 43.2$\pm$0.5\\
2458353.9356$\pm$0.00139  & 12.0$\pm$0.5 & 2458353.970725$\pm$0.001385  & 43.1$\pm$0.7\\
2458353.93838$\pm$0.00139  & 13.2$\pm$0.6 & 2458353.9735$\pm$0.00139  & 42.2$\pm$0.8\\
2458353.94116$\pm$0.00139  & 14.5$\pm$0.6 & 2458353.97628$\pm$0.00139  & 37.5$\pm$0.9\\
2458353.95128$\pm$0.00139  & 28.4$\pm$0.9 & 2458353.986905$\pm$0.001385  & 9.0$\pm$0.8\\
2458353.95406$\pm$0.00139  & 30.3$\pm$0.9 & 2458353.98968$\pm$0.00139  & 5.1$\pm$0.7\\
2458353.95684$\pm$0.00139  & 33.4$\pm$0.6 & 2458354.00357$\pm$0.0125 & 1.1$\pm$0.3\\
2458353.959615$\pm$0.001385  & 35.9$\pm$0.4 & - & -\\
\enddata
\tablenotetext{a}{The errors in the mean HJDs actually correspond to the length of the timerange over which the radio data were averaged}
\end{deluxetable*}

\startlongtable
\begin{deluxetable*}{cc|cc}
\tablecaption{Variation of flux density (total intensity) of HD 142990 in L band, data were acquired with the GMRT on 2014 November 8\label{tab:L_band_data}} 
\tablehead{
\hline
Mean  & Flux density        & Mean & Flux density\\
HJD\tablenotemark{a}  & mJy & HJD\tablenotemark{a} & mJy
}
\startdata
\hline
2456969.83808$\pm$0.001735 &  27.1$\pm$0.5 & 2456969.85892$\pm$0.001735 & 7.5$\pm$0.2\\
2456969.84156$\pm$0.001735 &  25.7$\pm$0.5 & 2456969.86958$\pm$0.001735 & 4.6$\pm$0.3\\
2456969.84502$\pm$0.001735 &  13.3$\pm$0.4 & 2456969.87306$\pm$0.001735 & 4.0$\pm$0.5\\ 
2456969.84850$\pm$0.001735 &  14.8$\pm$0.4 & 2456969.87654$\pm$0.001735 & 4.5$\pm$0.3\\
2456969.85198$\pm$0.001735 &  11.5$\pm$0.4 & 2456969.88522$\pm$0.006945 & 2.2$\pm$0.2\\
2456969.85544$\pm$0.001735 &   7.0$\pm$0.5 & 2456969.91157$\pm$0.01221  & 1.14$\pm$0.09\\
\hline
\enddata
\tablenotetext{a}{The errors in the mean HJDs actually correspond to the length of the timerange over which the radio data were averaged}
\end{deluxetable*}
\end{document}